\documentclass{article}

\input{tcilatex}

\begin{document}

\title{Classical Stability of Sudden \\
and Big Rip Singularities }
\author{John D. Barrow \thanks{%
J.D.Barrow@damtp.cam.ac.uk} $\ $ and Sean Z.W. Lip \thanks{%
S.Z.W.Lip@damtp.cam.ac.uk} \\
DAMTP, Centre for Mathematical Sciences,\\
Cambridge University, Wilberforce Road,\\
Cambridge CB3 0WA, United Kingdom $\ $}
\date{}
\maketitle

\begin{abstract}
We introduce a general characterization of sudden cosmological singularities
and investigate the classical stability of homogeneous and isotropic
cosmological solutions of all curvatures containing these singularities to
small scalar, vector, and tensor perturbations using gauge invariant
perturbation theory. We establish that sudden singularities at which the
scale factor, expansion rate, and density are finite are stable except for a
set of special parameter values. We also apply our analysis to the stability
of Big Rip singularities and find the conditions for their stability against
small scalar, vector, and tensor perturbations.

PACS: 98.80.-k
\end{abstract}

\begin{center}
\textbf{I. INTRODUCTION}
\end{center}

We have shown in \cite{jdb1} that general relativistic
Friedmann-Robertson-Walker (FRW) universes allow finite-time singularities
to occur in which the scale factor, $a(t)$, its time derivative, $\dot{a}$,
and the density, $\rho ,$ remain finite whilst a singularity occurs in the
fluid pressure, $p\rightarrow +\infty ,$ and the expansion acceleration,
with $\ddot{a}\rightarrow -\infty $. Remarkably, the strong energy condition 
$\rho +3p>0$ continues to hold. Analogous solutions are possible in which
the singularity can occur only in arbitrarily high derivatives of $a(t),$ 
\cite{jdb2}. This behaviour occurs independently of the 3-curvature of the
universe and can prevent closed FRW universes that obey the strong energy
condition from recollapsing \cite{jdb5}. These singularities can be seen
in a wider context by classifying the behaviors of FRW universes containing
matter with a pressure-density relation defined by $\rho +p=\gamma \rho
^{\lambda }$, as shown in \cite{jdb3}, and reviewed further in \cite{rev}.
The sudden singular behaviour found for a range of values of $(\gamma
,\lambda )$ also encompasses the evolution found in a number of simple bulk
viscous cosmologies studied in \cite{visc}. Subsequently, a number of
studies have been carried out which generalise these results to different
cosmologies, other density-pressure relationships, and theories of gravity 
\cite{jdb4, jdb4b}.

Other studies \cite{laz} have also related the sudden singularity behaviour
to the formal classifications of 'weak' singularities according to the
definitions introduced by Krolak \cite{krol} and Tipler \cite{tipl},
investigated the behaviour of geodesics, classified the other types of
future singularity that can arise during the expansion of the Universe \cite%
{odin}, discussed the role of different energy conditions on different
realizations of sudden singularities \cite{lake, jdb4}, and explored some
observational constraints on their possible future occurrence in our visible
Universe \cite{dab}. The Weyl invariant will not diverge on approach to a
sudden singularity (and there is no geodesic incompleteness \cite{geo}), so
it may represent part of a 'soft' future boundary of the universe with low
gravitational entropy -- which could be as close as 8.7 Myr in the future.

Most recently, the effects of quantum particle production have been studied
and have been found to leave sudden singularities in place \cite{fab}.
Specifically, it was shown that quantum particle production does not
dominate over the classical background density on approach to a sudden
singularity and does not stop it occurring or modify its properties, as can
be the case for the Big Rip future singularities \cite{qrip, no}. The
effects of loop quantum gravity have been studied in cosmologies exhibiting
classical sudden singularities and they may remove the sudden singularity
under certain particular conditions \cite{loop}. Sudden singularities have
also been studied due to their occurrence in various theories of modified
gravity {\cite{ANO}, and we would only expect these modifications to be
significant in this respect if they also dominate over general relativity
effects at late times}.

In this paper, we will extend previous studies by investigating the
classical stability of sudden singularities with respect to small
inhomogeneous scalar, vector, and tensor perturbations using the gauge
invariant formalism introduced by Mukhanov \cite{MUK2}. We introduce a new
characterization of sudden singularities in terms of the series expansion of
the expansion scale factor on approach to the singularity. We show that,
except for a subset of special parameter choices, they are stable to small
perturbations if and only if the density does not diverge near the
singularity. The latter is characteristic of sudden singularities. We also
extend this analysis and apply it to `Big Rip' singularities and determine
the conditions under which they are stable and unstable.

\begin{center}
\textbf{II. BACKGROUND THEORY}
\end{center}

Following {\cite{MUK1}}, we consider perturbations of a general FRW metric, 
\[
ds^{2}=a^{2}(\tau )\left( d\tau ^{2}-\delta _{ij}\left( 1+\frac{%
K(x^{2}+y^{2}+z^{2})}{4}\right) ^{-2}dx^{i}dx^{j}\right) , 
\]%
where $K=0,1,-1$ depending on whether the three-dimensional hypersurfaces of
constant $\tau $ time are spatially flat, closed or open. Here, $\tau $
denotes conformal time, and is related to the comoving proper time, $t,$ by $%
a\: d\tau =dt$. We shall also assume that the primary component of matter is a
perfect fluid with energy-momentum tensor (index notations run $1\leq
i,j\leq 3,0\leq \alpha ,\beta \leq 3$)

\begin{equation}
T_{\beta }^{\alpha }=(\rho +p)u^{\alpha }u_{\beta }-p\delta _{\beta
}^{\alpha },  \label{6}
\end{equation}%
but we shall not necessarily be specifying a specific equation of state
linking $p$ and $\rho $.

For this background universe, the equations for the scale factor $a(t)$,
density $\rho $ and pressure $p$ are as follows (in units with $8\pi G=1$
and $c=1$), where the overdot denotes $d/dt$: 
\begin{equation}
\dot{\rho}=-3\frac{\dot{a}}{a}(\rho +p)  \label{1}
\end{equation}%
\begin{equation}
\left( \frac{\dot{a}}{a}\right) ^{2}=\frac{\rho }{3}-\frac{K}{a^{2}}
\label{2}
\end{equation}%
\begin{equation}
\frac{\ddot{a}}{a}=-\left( \frac{\rho +3p}{6}\right)  \label{3}
\end{equation}%
and we shall assume an equation of state with a functional form $p=p(\rho )$%
. We can see by inspection that these equations permit finite-time
singularities such that $a,\dot{a}$ and $\rho $ are finite but $p$ and $%
\ddot{a}$ diverge as $t\rightarrow t_{s}$.

We shall adopt the following definition {\cite{CC}}: a sudden singularity
will be said to occur at time $t=t_{s}$ if the scale factor $a(t)$ can be
written in the form

\begin{equation}
a (t) = c_0 + c_1 (t_s - t)^{\lambda_1} + c_2 (t_s - t)^{\lambda_2} + \ldots
  \label{4}
\end{equation}%
in a generalised power series about $t_s$, where the $c_i, \lambda_i$ are
real constants, with $c_i \neq 0$, $c_0 > 0$ and $0 < \lambda_1 < \lambda_2
< \ldots $ with at least one of the $\lambda_i$ non-integral (so that some
derivative of $a (t)$ blows up near the singularity). Note that the series (%
\ref{4}) need not be infinite. Also, our form of (\ref{4}) assumes that the
sudden singularity occurs in the future; however, the stability results we
derive will also hold for past sudden singularities, defined in the
analogous way using

\begin{equation}
a (t) = c_0 + c_1 (t - t_s)^{\lambda_1} + c_2 (t - t_s)^{\lambda_2} + \ldots
  \label{5}
\end{equation}

This characterization encompasses the particular finite series expressions
for FRW models with sudden singularities introduced in Refs. \cite{jdb1,
jdb2}, those arising in the solutions found in Ref. \cite{jdb3}, and those
studied in Refs. \cite{rev, jdb4, jdb4b}.

The energy-momentum tensor defined in (\ref{6}) leads to the following
gauge-invariant perturbations {\cite{MUK2}}

\begin{equation}
\delta T_{0}^{0}=\delta \rho ,\delta T_{i}^{0}=\frac{1}{a_{{}}}(\rho
_{0}+p_{0})\delta u_{i},\delta T_{j}^{i}=-\delta p\delta _{j}^{i}=-p^{\prime
}(\rho )\delta \rho \delta _{j}^{i}.
\end{equation}

Here, $\delta \rho $, $\delta p$ and $\delta u_{i}$ are the gauge-invariant
perturbations of the density, pressure and velocity. We shall now consider
separately the behaviour of the tensor, vector and scalar modes as $%
t\rightarrow t_{s}.$

\begin{center}
\textbf{III. TENSOR PERTURBATIONS}
\end{center}

Under tensor perturbations, the most general form of the line element is 
\[
ds^{2}=a^{2}(\tau )\left( d\tau ^{2}-\left[ \left( 1+\frac{K}{4}%
(x^{2}+y^{2}+z^{2})\right) \delta _{ij}-h_{ij}\right] dx^{i}dx^{j}\right) , 
\]%
where $h_{i}^{i}=0$, $h_{ij}^{|j}=0$, where a slash indicates a covariant
derivative with respect to the spatial 3-metric. Note that the quantity $%
h_{ij}$ is gauge-invariant.

In conformal time, the equation for tensor perturbations is {\cite{BARDEEN}}

\begin{equation}
h_{ij}^{\prime \prime }+\frac{2a^{\prime }}{a}h_{ij}^{\prime }-\Delta
h_{ij}+2Kh_{ij}=0,
\end{equation}
where a prime denotes a derivative with respect to conformal time. The
cosmic time analogue of this equation is

\begin{equation}
a^2 \ddot{h}_{ij} + 3 a \dot{a} \dot{h}_{ij} + k^2 h_{ij} + 2 K h_{ij} = 0
\end{equation}
for a plane wave perturbation with wavenumber $k$. We now set $h_{i j} = v
e_{i j}$, where $e_{i j}$ is a time-independent polarisation tensor. This
leads to the differential equation 
\begin{equation}
a^2 \ddot{v} + 3 a \dot{a} \dot{v} + (k^2 + 2 K) v = 0 .  \label{TENSOR_OLD}
\end{equation}

From (\ref{4}), we see that in the limit $t \rightarrow t_s$, we have 
\begin{equation}
a (t) = c_0 + c_1 (t_s - t)^{\lambda_1} + \ldots 
\end{equation}
\begin{equation}
\dot{a} (t) = - c_1 \lambda_1 (t_s - t)^{\lambda_1 - 1} + \ldots 
\end{equation}
where $\lambda_1 > 0$, and $c_0, c_1 \neq 0$. We can substitute these into (%
\ref{TENSOR_OLD}) to obtain the asymptotic ordinary differential equation 
\begin{equation}
c_0^2 \ddot{v} - 3 c_0 c_1 \lambda_1 T^{\lambda_1 - 1} \dot{v} + (k^2 + 2 K)
v = 0,  \label{TENSOR}
\end{equation}
where we have taken $T = t_s - t$ and dots now indicate differentiation with
respect to $T$. We would like to investigate whether solutions to (\ref%
{TENSOR}) exhibit blow-up near $T = 0$.

In the case $\lambda_1 < 1$, a very similar analysis to that described in
the Appendix shows that $v$ tends to a constant as $T \rightarrow 0$.
Equations (\ref{66}) onwards still hold.

In the case $\lambda_1 \geq 1$, all coefficients of (\ref{TENSOR}) are
non-singular, so the ordinary differential equation is regular and hence has
no singularity at $T \rightarrow 0$. In fact, if $k^2 + 2 K > 0$, we end up
with a simple harmonic oscillator (which is damped if $\lambda_1 = 1$),
whose solutions are bounded at $T = 0$.

Hence, both long- and short-wavelength tensor perturbations are bounded near
the singularity. This clearly holds for all values of the $\lambda_i$, and
the sudden singularity is always stable against inhomogeneous
gravitational-wave perturbations.

\begin{center}
\textbf{IV. VECTOR PERTURBATIONS}
\end{center}

For vector perturbations, the most general form of the line element is 
\[
ds^{2}=a^{2}(\tau )\left( d\tau ^{2}+2S_{i}dx^{i}d\tau -\left[ \left( 1+%
\frac{K}{4}(x^{2}+y^{2}+z^{2})\right) \delta _{ij}-F_{i;j}-F_{j;i}\right]
dx^{i}dx^{j}\right) , 
\]%
where $S_{i}^{|i}=F_{i}^{|i}=0$. The quantity $V_{i}\equiv
S_{i}-F_{i}^{\prime }$ is gauge-invariant.

The only part of the energy-momentum tensor which contributes to vector
perturbations is {\cite{MUK2}}

\begin{equation}
\delta T^0_i = \frac{1}{a} (\rho_0 + p_0) \delta u_{\perp i},
\end{equation}
where $\delta u_{\perp i}$ is the part of $\delta u_{i}$ with zero
divergence, and $\rho _{0}$ and $p_{0}$ are the background density and
pressure, respectively. The equations for the vector perturbations are

\begin{equation}
(2 K - k^2) V_i = 2 a (\rho_0 + p_0) \delta u_{\perp i}
\end{equation}

\begin{equation}
a (V_{i, j} + V_{j, i})^. + 2 \dot{a} (V_{i, j} + V_{j, i}) = 0 .
\end{equation}

For both long- and short-wavelength perturbations, we obtain as a
consequence of the conservation of angular momentum:

\begin{equation}
V_i = const \times \frac{1}{a^2},
\end{equation}

\begin{equation}
\delta v_{}^i = const \times \frac{1}{a^4 (\rho_0 + p_0)},
\end{equation}
where the physical velocities $\delta v^{i}$ are defined by $\delta
v^{i}\equiv -a^{-1}\delta u_{\perp i}$. Hence, vector perturbations of the
metric are bounded on approach to the sudden singularity since $a\rightarrow
a_{s}<\infty ,\rho _{0}\rightarrow \rho _{s}<\infty $ and $p_{0}\rightarrow
\infty $.

\begin{center}
\textbf{V. SCALAR PERTURBATIONS}

\textbf{A. Overview}
\end{center}

We have shown, above, that tensor and vector perturbations do not diverge
near the sudden singularity, for all values of $\lambda _{i}$ in the form (%
\ref{4}). The analysis of scalar perturbations is slightly more involved,
and we shall need to consider various cases according to the values taken by 
$\lambda _{1}$ and $\lambda _{2}$. This is not unexpected. The sudden
singularity is primarily created by the behaviour of the pressure and so we
expect the scalar perturbation modes associated with pressure
inhomogeneities to play a significant role in controlling the stability.

Under scalar perturbations, the most general form of the line element is 
\[
ds^{2}=a^{2}(\tau )\left( (1+2\phi )d\tau ^{2}-2B_{;i}dx^{i}d\tau -\left[
\left( 1-2\psi \right) \left( 1+\frac{K}{4}(x^{2}+y^{2}+z^{2})\right) \delta
_{ij}+2E_{;ij}\right] dx^{i}dx^{j}\right) , 
\]%
and we can define the usual gauge-invariant quantities

\[
\Phi \equiv \phi -\frac{1}{a}\left[ a(B-E^{\prime })\right] ^{\prime } \text{%
and} \Psi \equiv \psi +\frac{a^{\prime }}{a}(B-E^{\prime }). 
\]

The equations for scalar perturbations are, following {\cite{MUK1}}:

\begin{equation}
\Phi = \Psi
\end{equation}

\begin{equation}
a^2 \ddot{\Phi} + (4 + 3 p^{\prime}(\rho)) a \dot{a}^{} \dot{\Phi} + (2 a 
\ddot{a} + ( \dot{a}^2 - K) (1 + 3 p^{\prime}(\rho)) + p^{\prime}(\rho) k^2)
\Phi = 0  \label{SCALAR}
\end{equation}
for plane wave perturbations with wavenumber $k$, where $p^{\prime }(\rho
)\equiv dp/d\rho $. We can then find the gauge-invariant perturbed
quantities as follows:

\begin{equation}
\delta \rho = - \frac{6 \dot{a}}{a} \dot{\Phi} - \left( \frac{2 k^2 + 6 \dot{%
a}^2}{a^2} \right) \Phi,
\end{equation}

\begin{equation}
\delta p=p^{\prime }(\rho )\delta \rho .
\end{equation}

\begin{center}
\textbf{B. The general case}
\end{center}
.
First, consider the case when $\lambda_1 \equiv \lambda$ is non-integral. In
this case we find, in the limit $t \rightarrow t_s$ :

\begin{equation}
a (t) = c_0 + c_1 (t_s - t)^{\lambda_{}} + \ldots 
\end{equation}

\begin{equation}
\dot{a} (t) = - c_1 \lambda_{} (t_s - t)^{\lambda - 1} + \ldots 
\end{equation}

\begin{equation}
\ddot{a}(t)=c_{1}\lambda _{{}}(\lambda -1)(t_{s}-t)^{\lambda _{{}}-2}+\ldots
\end{equation}%
and from (\ref{2}) and (\ref{3}) we obtain the leading-order approximations

\begin{equation}
\rho = \frac{3 K}{c_0^2} + \frac{3 \lambda^2 c_1^2}{c_0^2} (t_s - t)^{2
(\lambda - 1)} + \ldots 
\end{equation}

\begin{equation}
p = - \frac{K}{c_0^2} - \frac{2 c_1 \lambda (\lambda - 1)}{c_0} (t_s -
t)^{\lambda - 2} + \ldots 
\end{equation}
for the density and pressure, respectively. (Note that, regardless of the
value of $K$, the density diverges if and only if $\lambda < 1$, and the
pressure diverges if and only if $\lambda < 2$.) We also find that

\begin{equation}
p^{\prime}(\rho) = \frac{dp}{d \rho} = \frac{dp / dt}{d \rho / dt} = \frac{1%
}{3} \left( \frac{\dot{a}^3 - a^2 \ddot{a} \dot{} + K \dot{a}}{\dot{a} (a 
\ddot{a} - \dot{a}^2 - K)} \right) = - \frac{c_0 (\lambda - 2)}{3 c_1 \lambda%
} (t_s - t)^{- \lambda}
\end{equation}
to leading order, regardless of the value of $K$.

Substituting the above forms into (\ref{SCALAR}), and neglecting
higher-order terms, gives us the following equation (where $T = t_s - t$ and
dots now indicate differentiation with respect to $T$):

\begin{equation}
\ddot{\Phi} - (\lambda - 2) T^{- 1} \dot{\Phi} + \left( \frac{c_1}{c_0}
\lambda^2 T^{\lambda - 2} - \frac{(k^2 - 3 K) (\lambda - 2)}{3 c_0 c_1
\lambda} T^{- \lambda} - \frac{K}{c_0^2} \right) \Phi = 0  \label{S1}
\end{equation}
and we can neglect the last $( \frac{K\Phi }{c_{0}^{2}}) $ term since, for
all values of $\lambda $, it will be dominated in magnitude by one of the
other two terms making up the coefficient of $\Phi $, as $T\rightarrow 0$.
We will now consider the cases $\lambda <1$ and $\lambda >1$ separately.

If $\lambda <1,$ then (\ref{S1}) becomes

\begin{equation}
\ddot{\Phi} - (\lambda - 2) T^{- 1} \dot{\Phi} + \frac{c_1}{c_0} \lambda^2
T^{\lambda - 2} \Phi = 0
\end{equation}
and applying the substitutions $P = T^{(1 - \lambda) / 2} \Phi$ and $x =
CT^{\lambda / 2}$ gives the equation

\begin{equation}
P^{\prime\prime}x^2 + P^{\prime}x + \left( \frac{4 c_1 x^2}{c_0 C^2} - \frac{%
(\lambda - 1)^2}{\lambda^2} \right) P =3 0
\end{equation}
where primes denote differentiation with respect to $x$. If $c_1 > 0$, we
set $C = 2 \sqrt{c_1/c_0}$, and obtain a Bessel equation with solution

\bigskip 
\[
P(x)=\tilde{A}J_{\nu }(x)+\tilde{B}Y_{\nu }(x) 
\]
.
\[
\nu =\frac{\lambda -1}{\lambda } 
\]
with $\tilde{A},\tilde{B}$ arbitrary constants. Since we are considering the
limit $x\rightarrow 0$, the leading-order solution is

\begin{equation}
P (x) = A x^{\frac{\lambda - 1}{\lambda}} + B x^{\frac{1 - \lambda}{\lambda}}
\end{equation}
or

\begin{equation}
\Phi (t) = A (t_s - t)^{\lambda - 1} + B
\end{equation}
where $A, B$ are new constants. Therefore, since $\lambda <1$, $\Phi (t)$
diverges in general as $t\rightarrow t_{s}$. The case $c_1 < 0$ can be
treated similarly. Note that this result holds, independently of $k$, for
both long- and short-wavelength perturbations and for all $K$.

Now suppose that $\lambda > 1$. Then (\ref{S1}) becomes

\begin{equation}
\ddot{\Phi} - (\lambda - 2) T^{- 1} \dot{\Phi} - D T^{- \lambda} \Phi = 0
\end{equation}
where $D=\frac{(k^{2}-3K)(\lambda -2)}{3c_{0}c_{1}\lambda }$. Analogously to
the above, we substitute $P=T^{(1-\lambda )/2}\Phi $, and then $x=C
T^{1-\lambda /2}$, which leads to the Bessel equation

\begin{equation}
P^{\prime\prime}x^2 + P^{\prime}x + \left( x^2 - \frac{(1 - \lambda)^2}{(2 -
\lambda)^2} \right) P = 0
\end{equation}
\[
\nu =\frac{\lambda -1}{2 -\lambda } 
\]
where we have set $C = 2 \sqrt{|D|}/|\lambda - 2|$.

As $x\rightarrow 0,$ this has the asymptotic solution

\begin{equation}
P (x) = A x^{\frac{\lambda - 1}{2 - \lambda}} + B x^{\frac{1 - \lambda}{2 -
\lambda}}
\end{equation}

or

\begin{equation}
\Phi (t) = A (t_s - t)^{\lambda - 1} + B.
\end{equation}

Thus, if $\lambda >1$, the scalar metric perturbations do not diverge as $%
t\rightarrow t_{s}$. Again, note that this holds for all values of $K$, and
for both long- and short-wavelength perturbations.

Hence, the scalar metric perturbations diverge for all $0<\lambda _{1}<1$,
and are bounded for all non-integral $\lambda _{1}>1$. In fact, the analysis
above also holds for all integral $\lambda _{1}>2$. In this case, the
density and pressure tend to constants near the singularity, and the metric
perturbations do not diverge. Thus, we just need to deal separately with the
boundary cases of $\lambda _{1}=1$ and $\lambda _{1}=2$.

\begin{center}
\textbf{C. The case $\lambda_1 = 1, \lambda_2 \neq 2$}
\end{center}

Consider the case $\lambda _{1}=1<\lambda _{2}<\ldots $ and assume first
that $\lambda _{2}\neq 2.$ So, as $t\rightarrow t_{s}$ :

\begin{equation}
a (t) = c_0 + c_1 (t_s - t) + c_2 (t_s - t)^{\lambda_2} + \ldots 
\end{equation}

\begin{equation}
\dot{a} (t) = - c_1 - c_2 \lambda_2 (t_s - t)^{\lambda_2 - 1} + \ldots 
\end{equation}
\begin{equation}
\ddot{a} (t) = c_2 \lambda_2 (\lambda_2 - 1) (t_s - t)^{\lambda_2 - 2} +
\ldots 
\end{equation}

\begin{equation}
\ddot{a} \dot{} (t) = - c_2 \lambda_2 (\lambda_2 - 1) (\lambda_2 - 2) (t_s -
t)^{\lambda_2 - 3} + \ldots 
\end{equation}

\begin{equation}
\rho = \frac{3 (K + c_1^2)}{c_0^2} + \ldots 
\end{equation}
\begin{equation}
p = - \frac{(K + c^2_1)}{c^2_0} - \frac{2 c_2 \lambda_2 (\lambda_2 - 1) (t_s
- t)^{\lambda_2 - 2}}{c_0} + \ldots 
\end{equation}

We can see that $\rho$ tends to a constant as we approach the singularity,
but $p$ tends to a constant if and only if $\lambda_2 \geq 2$, and diverges
otherwise.

We can also obtain the following expressions: 
\[
p^{\prime}(\rho) = \left\{ 
\begin{array}{lll}
- \frac{c_0 (\lambda_2 - 2)}{3 c_1} (t_s - t)^{- 1} & \text{if} & 1 <
\lambda_2 < 2 \\ 
\frac{c_0^2 c_2 \lambda_2 (\lambda_2 - 1) (\lambda_2 - 2)}{3 c_1 (K + c_1^2)}
(t_s - t)^{\lambda_2 - 3} & \text{if} & 2 < \lambda_2 < 3 \\ 
\frac{6 c_0^2 c_2 - c_1^3 - Kc_1}{3 c_1 (K + c_1^2)} & \text{if} & \lambda_2
= 3 \\ 
- \frac{1}{3} & \text{if} & \lambda_2 > 3%
\end{array}
\right. 
\]
if $K + c_1^2 \neq 0$, and 
\[
p^{\prime}(\rho) = - \frac{c_0 (\lambda_2 - 2)}{3 c_1} (t_s - t)^{- 1} 
\]
for all $\lambda_2$, if $K + c_1^2 = 0$ (i.e. $K = - 1,$ $c_1 = \pm 1$).

First, we consider the case $K+c_{1}^{2}\neq 0$. Substituting the forms of $%
p^{\prime }(\rho )$ into (\ref{SCALAR}) gives 
\[
\begin{array}{lll}
\ddot{\Phi}+AT^{-1}\dot{\Phi}+BT^{-1}\Phi =0 & \text{if} & 1<\lambda _{2}<2
\\ 
\ddot{\Phi}+AT^{\lambda _{2}-3}\dot{\Phi}+BT^{\lambda _{2}-3}\Phi =0 & \text{%
if} & 2<\lambda _{2}<3 \\ 
\ddot{\Phi}+A\dot{\Phi}+B\Phi =0 & \text{if}. & \lambda _{2}\geq 3%
\end{array}%
\]%
where $A$ and $B$ are always constants. The first differential equation
yields an asymptotic solution $\Phi (t)\approx A^{\prime }+B^{\prime
}(t_{s}-t)^{\lambda _{2}-1},$ so the scalar perturbations are bounded in
this case. Showing that solutions of the second differential equation are
bounded as $T\rightarrow 0$ is more cumbersome: a derivation is given in the
Appendix. For $\lambda _{2}\geq 3$, the coefficients of the differential
equation are constants, so the general solution has the form

\begin{equation}
\Phi (t) \approx A^{\prime}\exp (\Lambda_1 (t_s - t)) + B^{\prime}\exp
(\Lambda_2 (t_s - t)),
\end{equation}
which is bounded as $t \rightarrow t_s$.

The same analysis can be used to show that the solution in the case $%
K+c_{1}^{2}=0$ is also bounded as $T\rightarrow 0$. Notice that these results hold for
both long- and short-wavelength perturbations.

\begin{center}\textit{\ }
\textbf{D. The case $\lambda_1 = 1, \lambda_2 = 2$}
\end{center}

Now let $\lambda_1 = 1, \lambda_2 = 2 < \lambda_3 < \ldots $ As $t
\rightarrow t_s$ :

\begin{equation}
a (t) = c_0 + c_1 (t_s - t) + c_2 (t_s - t)^2 + c_3 (t_s - t)^{\lambda_3} +
\ldots 
\end{equation}

\begin{equation}
\dot{a} (t) = - c_1 - 2 c_2 (t_s - t)^{} - c_3 \lambda_3 (t_s -
t)^{\lambda_3 - 1} \ldots 
\end{equation}
\begin{equation}
\ddot{a} (t) = 2 c_2 + c_3 \lambda_3 (\lambda_3 - 1) (t_s - t)^{\lambda_3 -
2} + \ldots 
\end{equation}
\begin{equation}
\ddot{a} \dot{} (t) = - c_3 \lambda_3 (\lambda_3 - 1) (\lambda_3 - 2) (t_s -
t)^{\lambda_3 - 3} + \ldots 
\end{equation}

\begin{equation}
\rho = \frac{3 (K + c_1^2)}{c_0^2} + \ldots 
\end{equation}

\begin{equation}
p = - \frac{(K + c^2_1)}{c^2_0} - \frac{4 c_2}{c_0} + \ldots 
\end{equation}

so the density and pressure tend to constants near the singularity. Also: 
\[
p^{\prime }(\rho )=\left\{ 
\begin{array}{lll}
\frac{c_{0}^{2}c_{3}\lambda _{3}(\lambda _{3}-1)(\lambda _{3}-2)}{%
3c_{1}(c_{1}^{2}+K-2c_{0}c_{2})}(t_{s}-t)^{\lambda _{3}-3} & \text{if} & 
2<\lambda _{3}<3 \\ 
\frac{6c_{0}^{2}c_{3}-Kc_{1}-c_{1}^{3}}{3c_{1}(c_{1}^{2}+K-2c_{0}c_{2})} & 
\text{if} & \lambda _{3}=3 \\ 
\frac{c_{1}^{2}+K}{3(2c_{0}c_{2}-c_{1}^{2}-K)} & \text{if} & \lambda _{3}>3%
\end{array}%
\right. 
\]%
and we can now substitute these into (\ref{SCALAR}) to get the following
functional forms of the scalar perturbation equation (where $T=t_{s}-t$,
dots indicate derivatives with respect to $T$, and $A,B$ are constants): 
\[
\begin{array}{lll}
\ddot{\Phi}+AT^{\lambda _{3}-3}\dot{\Phi}+BT^{\lambda _{3}-3}\Phi =0 & \text{%
if} & 2<\lambda _{3}<3 \\ 
\ddot{\Phi}+A\dot{\Phi}+B\Phi =0 & \text{if} & \lambda _{3}\geq 3%
\end{array}%
\]%
Equations of this form arose in the previous section, and we showed that
their solutions are bounded as $T\rightarrow 0$.

The above expressions for $p^{\prime }(\rho )$ are not well-defined if $%
K=2c_{0}c_{2}-c_{1}^{2}$. For this special case, we can calculate 
\[
p^{\prime }(\rho )=\left\{ 
\begin{array}{lll}
-\frac{c_{0}(\lambda _{3}-2)}{3c_{1}}(t_{s}-t)^{-1} & \text{if} & 2<\lambda
_{3}<3 \\ 
-\frac{c_{0}}{3c_{1}}(t_{s}-t)^{-1} & \text{if} & \lambda _{3}\geq 3%
\end{array}%
\right. 
\]%
and the corresponding scalar perturbation equation is

\[
\ddot{\Phi}+A\dot{\Phi}+B\Phi /T=0. 
\]%
Solutions of this differential equation can be expressed in terms of
Whittaker functions, and are bounded as $T\rightarrow 0$.

Hence, the scalar perturbations are bounded for $\lambda _{1}=1$, $\lambda
_{2}=2$. This is quite a strong result, since most `nice' functions $a(t)$
can be expressed as power series. We have shown that for all such functions,
as long as the coefficient of the $t$ term is non-zero near the singularity
(i.e. $\dot{a}(t_{s})\neq 0$), the sudden singularity is a stable solution
of the FRW equations.

\begin{center}
\textbf{E. The case $\lambda_1 = 2$}
\end{center}

The final case to consider is:

\begin{equation}
a (t) = c_0 + c_1 (t_s - t)^2 + c_2 (t_s - t)^{\lambda_2} + \ldots 
\end{equation}
where the $c_{i}\neq 0$, $c_{0}>0$ and $2<\lambda _{2}<\ldots $ with at
least one of the $\lambda _{i}$ non-integral. As $t\rightarrow t_{s}$, we
have:

\begin{equation}
\dot{a} (t) = - 2 c_1 (t_s - t) - c_2 \lambda_2 (t_s - t)^{\lambda_2 - 1} +
\ldots 
\end{equation}
\begin{equation}
\ddot{a} (t) = 2 c_1 + c_2 \lambda_2 (\lambda_2 - 1) (t_s - t)^{\lambda_2 -
2} + \ldots 
\end{equation}
\begin{equation}
\ddot{a} \dot{} (t) = - c_2 \lambda_2 (\lambda_2 - 1) (\lambda_2 - 2) (t_s -
t)^{\lambda_2 - 3} + \ldots 
\end{equation}
and $\rho \rightarrow 3 K / c_0^2, p \rightarrow - (4 c_1 c_0 + K) / c^2_0$.

First, let us assume $K=0$. There are three cases to consider for $p^{\prime
}(\rho )$: 
\[
p^{\prime }(\rho )=\left\{ 
\begin{array}{lll}
-\frac{c_{0}c_{2}\lambda _{2}(\lambda _{2}-1)(\lambda _{2}-2)}{12c_{1}^{2}}%
(t_{s}-t)^{\lambda _{2}-4} & \text{if} & 2<\lambda _{2}<6 \\ 
\frac{2c_{1}^{3}-30c_{0}^{2}c_{2}}{3c_{0}c_{1}^{2}}(t_{s}-t)^{2} & \text{if}
& \lambda _{2}=6 \\ 
\frac{2c_{1}}{3c_{0}}(t_{s}-t)^{2} & \text{if} & \lambda _{2}>6%
\end{array}%
\right. 
\]%
We now substitute these expressions into the equation for scalar
perturbations (\ref{SCALAR}). Using $A,B,\ldots $ to denote constants, the
functional forms of the differential equations obtained are: 
\[
\begin{array}{lll}
\ddot{\Phi}+AT^{\lambda _{2}-3}\dot{\Phi}+(BT^{\lambda
_{2}-2}+Ck^{2}T^{\lambda _{2}-4})\Phi =0 & \text{if} & 2<\lambda _{2}<4 \\ 
\ddot{\Phi}+AT\dot{\Phi}+B\Phi =0 & \text{if} & \lambda _{2}\geq 4%
\end{array}%
\]%
\emph{\ \ \ \ \ }The solutions of the second differential equation are
bounded as $T\rightarrow 0$. If we assume that $k\neq 0$, the solution of
the first differential equation can be expressed in terms of hypergeometric
functions, and is bounded as $T\rightarrow 0$.

Now suppose that $K \neq 0$. There are two sub-cases: $K \neq 2 c_1 c_0$ and 
$K = 2 c_1 c_0$. For the first sub-case, we have: 
\[
p^{\prime}(\rho) = \left\{ 
\begin{array}{lll}
\frac{c_0^2 c_2 \lambda_2 (\lambda_2 - 1) (\lambda_2 - 2)}{6 c_1 (K - 2 c_1
c_0)} (t_s - t)^{\lambda_2 - 4} & \text{if} & 2 < \lambda_2 < 4 \\ 
\frac{24 c_0^2 c_2 - 2 K c_1}{6 c_1 (K - 2 c_1 c_0)} & \text{if} & \lambda_2
= 4 \\ 
\frac{K}{3 (2 c_1 c_0 - K)} & \text{if} & \lambda_2 > 4%
\end{array}
\right. 
\]
and the corresponding differential equations are 
\[
\begin{array}{lll}
\ddot{\Phi} + A T^{\lambda_2 - 3} \dot{\Phi} + (B T^{\lambda_2 - 2} + C k^2
T^{\lambda_2 - 4}) \Phi = 0 & \text{if} & 2 < \lambda_2 < 4 \\ 
\ddot{\Phi} + A T \dot{\Phi} + B \Phi = 0 & \text{if} & \lambda_2 \geq 4%
\end{array}
\]
with the same stability results as above.

Finally, we treat the special case $K=2c_{1}c_{0}$. We obtain: 
\[
p^{\prime }(\rho )=\left\{ 
\begin{array}{lll}
-\frac{c_{0}(\lambda _{2}-2)}{6c_{1}}(t_{s}-t)^{-2} & \text{if} & 2<\lambda
_{2}<4 \\ 
-\frac{c_{0}}{3c_{1}}(t_{s}-t)^{-2} & \text{if} & \lambda _{2}\geq 4%
\end{array}%
\right. 
\]%
and this corresponds to: 
\[
\text{$\ddot{\Phi}-\frac{6c_{1}C}{c_{0}}T^{-1}\dot{\Phi}+\frac{3(k^{2}-K)C}{%
c_{0}^{2}}T^{-2}\Phi =0$} 
\]%
for all $\lambda _{2}>2$, where $p^{\prime }(\rho )=CT^{-2}$. In general,
the solutions of this equation are of the form $\Phi (T)=T^{\gamma }$ where

\[
\gamma ^{2}+(-6c_{1}C/c_{0}-1)\gamma +3(k^{2}-K)C/c_{0}^{2}=0. 
\]

For boundedness we need the two roots of this equation to be non-negative,
i.e. we require $-6c_{1}C\leq c_{0}$ and $(k^{2}-K)C\geq 0$. It is easy to
see that the first condition only holds for $\lambda _{2}\leq 3$. So, in
this sub-case, we only have boundedness if $\lambda _{2}\leq 3$ and $%
(k^{2}-2c_{1}c_{0})C\geq 0$, and divergence otherwise.

So all the perturbations are bounded except for the very special case $K
\neq 0, K = 2 c_1 c_0, \lambda_2 > 3$ (and perhaps $\lambda_2 \leq 3$).

\begin{center}
\textbf{F. Consideration of the sign of $c_{s}^2$}
\end{center}

Note that our analysis here has not assumed a simple fluid equation of state of the
form $p = w\rho$, or indeed any functional relation between the density and pressure 
of the matter source, as would characterize a perfect fluid of k-essence or its generalizations
(see \cite{BERTACCA} for a discussion). Thus there is no general constraint arising from the
positivity of the square of an effective speed of sound, $c_{s}^{2}$. In cases that reduce
to fluids, or to k-essence and its relatives, it should be possible to introduce further
constraints upon the series coefficients in order to preserve the positivity of $c_{s}^{2}$.

\begin{center}
\textbf{VI. EXTENSIONS TO BIG RIP MODELS}
\end{center}

We can extend the above analysis easily to Big Rip models \cite{rip}, where
the scale factor behaves as:

\begin{equation}
a (t) = c_0 (t_s - t)^{\eta_0} + c_1 (t_s - t)^{\eta_1} + \ldots 
\end{equation}
where $\eta_0 < 0, \eta_0 < \eta_1 < \ldots$, and also $c_0 > 0$, $c_i \neq
0 $ for all $i$. Then, to leading order, we have

\begin{equation}
\dot{a} (t) = - \eta_0 c_0 (t_s - t)^{\eta_0 - 1} + \ldots 
\end{equation}
\begin{equation}
\ddot{a} (t) = \eta_0 (\eta_0 - 1) c_0 (t_s - t)^{\eta_0 - 2} + \ldots 
\end{equation}
\begin{equation}
\ddot{a} \dot{} (t) = - \eta_0 (\eta_0 - 1) (\eta_0 - 2) c_0 (t_s -
t)^{\eta_0 - 3} + \ldots 
\end{equation}

\begin{equation}
\rho = 3 \eta_0^2 (t_s - t)^{- 2} + \ldots 
\end{equation}
\begin{equation}
p = (2 - 3 \eta_0) \eta_0 (t_s - t)^{- 2} + \ldots 
\end{equation}

\begin{equation}
p^{\prime}(\rho) = 2 / 3 \eta_0 - 1 + \ldots 
\end{equation}

Note that $a (t), \rho, p \rightarrow \infty$ as $t \rightarrow t_s$. It is
easy to see that, near the singularity, the tensor perturbations $h_{ij}$
and the vector perturbations $V_i$ decay to zero. However, the physical
velocities $\delta v^i$ are proportional to $(t_s - t)^{- 2 - 4 \eta_0}$, so
for them to be finite as $t \rightarrow t_s$ we need $\eta_0 \leq - 1 / 2.$

Scalar perturbations obey the following equation:

\begin{equation}
\ddot{\Phi} + (2 + \eta_0) \frac{\dot{\Phi}}{T} + \left( \frac{2 (K - k^2)}{%
c_0^2} + \frac{2 (k^2 - 3 K)}{3 \eta_0 c_0^2} \right) T^{- 2 \eta_0} \Phi = 0
\label{BIGRIP}
\end{equation}
where $T = t_s - t$ and dots denote differentiation with respect to $T$. A
solution of (\ref{BIGRIP}) can be found in terms of Bessel functions, and we
find that the solutions asymptotically tend to

\begin{equation}
\Phi (T) = C_1 + C_2 T^{- 1 - \eta_0} .
\end{equation}

So there is divergence as $t\rightarrow t_{s}$ if $\eta _{0}>-1$, otherwise $%
\Phi $ is bounded.

Note that these results match those of {\cite{FJG}}, which shows that, for
equations of state of the form $p = \alpha \rho$ (where $\alpha$ is a
constant), there is a discrepancy in behaviour between the cases $\alpha > - 5 /
3 $ and $\alpha < - 5 / 3$. It can easily be checked that these correspond
to the cases $\eta_0 < - 1$ and $\eta_0 > - 1$ respectively.

\begin{center}
\textbf{VII. CONCLUSION}
\end{center}

We have produced a simple general characterization of sudden singularities
in FRW universes. By the use of gauge invariant perturbation theory we have
investigated whether the existence of sudden singularities in a FRW
cosmology is stable to small scalar, vector, and tensor inhomogeneities. We
have shown that the existence of sudden singularities is stable when the
density is bounded near the singularity except for some cases with special
parameter choices. This result holds regardless of whether the background
metric is spatially flat, closed or open. We also applied our analysis to a
complementary characterization of Big Rip singularities and showed that they
are stable if the leading term in the time-dependence of the FRW scale
factor is proportional to $a(t)=(t_{s}-t)^{\eta _{0}}$ where $\eta _{0}\leq
-1$, and unstable otherwise. As discussed in the introduction, there have
been a number of identifications of sudden singularity occurrence in
theories of gravity other than general relativity. The approach described in
this paper can also be straightforwardly applied in these theories to
determine whether the sudden singularities that occur there are also stable.

\begin{center}
\textbf{ACKNOWLEDGMENTS}
\end{center}

S. Lip is supported by the Gates Cambridge Trust. The authors would like to
thank Anthony Ashton, J\'{u}lio Fabris and Alain Goriely for discussions.

\begin{center}
\textbf{APPENDIX}

\textbf{Equations of the form $\ddot{\Phi} + AT^s \dot{\Phi} + BT^s \Phi = 0$%
}
\end{center}

Consider the equation

\begin{equation}
\ddot{\Phi} + AT^s \dot{\Phi} + BT^s \Phi = 0
\end{equation}
where $\Phi = \Phi (T)$, $- 1 < s < 0$ and $A, B$ are arbitrary constants.
We want to investigate the behaviour of this equation as $T \rightarrow 0$.

We first reduce the equation to canonical form $\ddot{\phi} + f (T) \phi = 0$%
. The substitution $\Phi = \phi \exp (- AT^{s + 1} / 2 (s + 1))$ achieves
this, and yields

\begin{equation}
\ddot{\phi} + \left( BT^s - \frac{1}{2} AsT^{s - 1} - \frac{1}{4} A^2 T^{2
s} \right) \phi = 0 .
\end{equation}

We now let $Y (t) = t \phi (1 / t)$ and this gives

\begin{equation}
\ddot{Y} + \left( Bt^{- s - 4} - \frac{1}{2} Ast^{- s - 3} - \frac{1}{4} A^2
t^{- 2 s - 4} \right) Y = 0 .
\end{equation}
where $t=1/T$ and we now study the behaviour as $t\rightarrow +\infty $. The
substitutions $Y(t)=\exp (\phi (t))$ and $u(t)=\dot{\phi}(t)$ transform (83)
into a Riccati equation, which (after neglecting sub-leading terms) becomes:

\begin{equation}
\dot{u} + u^2 - \frac{1}{2} Ast^{- s - 3} = 0 .  \label{66}
\end{equation}

Finally we make the substitution $u = \dot{y} / y$ to obtain the
Emden-Fowler equation

\begin{equation}
\ddot{y} = \frac{1}{2} Asyt^{- s - 3} .
\end{equation}

This has a solution in terms of Bessel functions:

\begin{equation}
y (t) = t^{1 / 2} \left( C_1 J_{\frac{1}{s + 1}} \left( \frac{2 \sqrt{- D}}{%
- s - 1} t^{(- s - 1) / 2} \right) + C_2 Y_{\frac{1}{s + 1}} \left( \frac{2 
\sqrt{- D}}{- s - 1} t^{(- s - 1) / 2} \right) \right) \text{ if } D < 0
\end{equation}
\begin{equation}
y (t) = t^{1 / 2} \left( C_1 I_{\frac{1}{s + 1}} \left( \frac{\sqrt{D}}{- s
- 1} t^{(- s - 1) / 2} \right) + C_2 K_{\frac{1}{s + 1}} \left( \frac{\sqrt{D%
}}{- s - 1} t^{(- s - 1) / 2} \right) \right) \text{ if } D > 0
\end{equation}
where $D = As / 2$. We use the asymptotic behaviour of the Bessel functions
as $t \rightarrow + \infty$ to obtain

\begin{equation}
y (t) \sim C_3 + C_4 t
\end{equation}
and $u(t)\sim 1/t$. So $Y(t)\sim t$ and $\Phi (T)\rightarrow const$ as $%
T\rightarrow 0$.

\end{document}